%%AmSTeX

\documentstyle{amsppt} \magnification=\magstep1 
\hsize=6.5 truein \hcorrection{.375in} \vsize=8.5 truein 
\parindent=20pt \baselineskip=12pt \TagsOnRight
\NoBlackBoxes\footline{\hss\tenrm\folio\hss}

%%AmSTeX

\centerline{The Radial Part of the Zero-Mode Hamiltonian}
\centerline{ for Sigma Models with Group Target Space}

\bigskip\bigskip

\centerline{Doug Pickrell}
\centerline{Mathematics Department}
\centerline{University of Arizona}
\centerline{Tucson, AZ 85721}
\centerline{pickrell\@math.arizona.edu}

\bigskip\bigskip

\flushpar Abstract:  In this note we use geometric 
arguments to derive a possible form for the radial part 
of the ``zero-mode Hamiltonian'' for the two dimensional 
sigma model with target space $S^3$, or more generally a 
compact simply connected Lie group.  

\bigskip

\flushpar Mathematics Subject Classifications (2000):  
81T40, 22E67, 58D20.  

\smallskip

\flushpar Key Words:  Sigma model, Hamiltonian, loop 
group, Wiener measure.  

\bigskip\bigskip

\centerline{\S 0. Introduction.}

\bigskip

At the classical level a two dimensional sigma model 
with arbitrary Riemannian target space is conformally 
invariant (see \S 3 of [Ga], or \S 1 below, for the meaning of 
these terms).  At the quantum level (assuming the 
theory makes sense), it is believed that conformal 
invariance is broken whenever the Ricci curvature of 
the target is nonvanishing (see \S 3 of [Ga]).  This is the 
case for the targets considered in this paper, compact 
simply connected Lie groups with biinvariant metric.  

For a given quantum 2D sigma model with compact 
target space, one expects that for each finite radius $R$, 
there is a Hilbert space of states, $H(S^1_R)$, associated to 
the circle with radius $R$, and there is a $\underline {trace}$ $\underline {
class}$ 
operator, $U(\Sigma )$, from incoming to outgoing state spaces, 
associated to each oriented compact Riemannian surface 
$\Sigma$, with parameterized geodesic boundary components, 
such that sewing of surfaces corresponds to composition 
of operators.  The sewing property expresses the 
locality of quantum field theory, as interpreted by Segal; 
the trace class condition arises from the expectation 
that finite numbers can be associated to the path 
integrals corresponding to closed surfaces (see \S 2.6 of 
[Ga]).  In particular the infinitesimal generator for the 
one parameter family $U(S^1_R\times [0,t])$, the Hamiltonian $H_R$, 
should have a discrete spectrum.  

In the case of a conformally invariant quantum sigma 
model, the Hilbert space $H(S^1_R)$ and Hamiltonian $H_R$ are 
essentially independent of $R$.  Otherwise there is 
dependence on $R$, and for the models we consider, it is 
expected that in the limit $R\uparrow\infty$, the Hamiltonian $H=
H_{\infty}$ 
has a mass gap and continuous spectrum.  

In this paper we will assume that the Hilbert spaces 
$H(S^1_R)$ have a certain concrete mathematical form.  We 
will also introduce some simplifying assumptions 
regarding the action of the Hamiltonian operators $H_R$.  
On the basis of these assumptions, we will draw some 
conclusions about the form of the radial part of the 
``zero-modes'' of these operators.  
  
The arguments in this paper are most complete in the 
limiting case $R\uparrow\infty$.  We assume that in this limit the 
space of states of the 2D sigma model, with simply 
connected group target space $K$, has the form 
$\Cal H(S^1_{\infty})=L^2(\mu )$, where $\mu$ is a certain canonical measure on 
a distribution-like completion of the loop space $LK$.  We 
think of this measure as specifying the vacuum for the 
theory.  This is motivated by the known form for the 
vacuum of the conformally invariant WZW model (see 
\S 4.1 of [Ga]).  We also assume that the Hamiltonian acts 
as a second order differential operator on a certain 
subspace of zero-mode states.  In the case of $S^3$, on the 
basis of these assumptions, we find that the radial part 
of this ``zero-mode Hamiltonian'' is equivalent to 
$$-(\frac d{dr})^2+\frac 14-\frac {15}4sech^2(r),\tag 0.1$$
(acting on odd functions) up to a scale factor (the mass 
parameter).  This operator has a unique ground state and 
a mass gap.  The spectrum of this radial part does not 
reflect expected features of the full spectrum, such as 
jumps in multiplicity corresponding to multiparticle 
states.  There is a mechanism which should produce 
these jumps in the spectrum of the full zero-mode 
Hamiltonian, namely the discreteness of the $K\times K$ 
isotypic decomposition.  Unfortunately the zero-mode 
Hamiltonian cannot be determined on mathematical 
grounds alone from its radial part.  

In \S 1 I will more fully explain the motivation for the 
conjecture.  This involves a brief review of the 
Hamiltonian formalism for the two dimensional sigma 
model.  

In \S 2 I will briefly review some of the mathematics 
which \S 1 depends upon, especially the construction of the 
measure $\mu$ (which is carried out in [Pi]).  This involves 
understanding the limit of Wiener measure on $LK$, in 
terms of Riemann-Hilbert factorization, as inverse 
temperature tends to zero (which corresponds to $R\to\infty$).  
For the purposes of this paper, the key is Step 7 in \S 2, 
which is a conjectured formula for the spherical 
transform of the diagonal (or zero-mode) distribution of 
the measure $\mu$, in terms of an affine analogue of 
Harish-Chandra's famous {\bf c}-function (the mathematical 
conclusions of this paper depend exclusively on this 
affine {\bf c}-function, and many readers may care to skip 
the infinite dimensional measure-theoretic considerations; 
however, it should be borne in mind that the fact this 
function arises as a transform for a measure on 
(distributional) loop space is a key argument for the 
plausibility of our physical claims).  

In \S 3 I will present the calculations which lead to $(0.1)$, 
and to an analogous conjecture for the massive 
deformation of the conformally invariant WZW model.  

In $\S 4$ I will discuss the case of finite radius.  The 
mathematical underpinnings for this section are not as 
strong as in the case $R\uparrow\infty$, because the analogue of the 
measure $\mu$ has not been rigorously constructed.  
However there does appear to be a natural conjecture 
for the zero-mode distribution.  This involves 
understanding a quantum (theta function) deformation of 
the affine {\bf c}-function.  Most of this section revolves 
around some nontrivial positivity checks for this 
deformation.

\bigskip

\centerline{\S 1. Origins of the Conjecture.}

\bigskip

We initially suppose that space is the circle, $S^1$, so that 
spacetime is $\Sigma =S^1\times \Bbb R$, with coordinates $(\theta 
,t)$.  We also 
initially suppose that the target space is $X$, an arbitrary 
Riemannian manifold.  

The classical fields for the sigma model with target 
space $X$ are maps $x:\Sigma\to X$, and the action is the kinetic 
energy function 
$$\Cal A(x)=\frac 12\int_{\Sigma}\langle dx\wedge *dx\rangle =\frac 
12\int_{\Sigma}\{\vert\frac {\partial x}{\partial t}\vert^2+\vert\frac {
\partial x}{\partial\theta}\vert^2\}d\theta dt.\tag 1.1$$
This action is conformally invariant, meaning that if the 
metric $ds$ of $\Sigma$ is changed to $\rho ds$, where $\rho$ is a positive 
function, the action remains unchanged.  In particular 
the action depends upon the radius of the circle and 
time scale in a covariant way.  

The time zero fields constitute the loop space 
$LX=Map(S^1,X)$ (when it is useful to denote the degree 
of smoothness, we will use a subscript, e.g.  $L_{C^0}X$ will 
denote the manifold of continuous loops).  The tangent 
space to $LX$ at $x$ is naturally identified with $\Omega^0(x^{*}
TX)$, 
the space of vector fields along the loop $x$.  There is a 
Riemannian metric on this tangent space, given by 
$$\langle v,w\rangle_x=\int_{S^1}\langle v(\theta ),w(\theta )\rangle_{
x(\theta )}d\theta ,\tag 1.2$$
where $v(\theta ),w(\theta )\in TX\vert_{x(\theta )}$, and $\langle
\cdot ,\cdot\rangle_{x(\theta )}$ denotes the inner 
product (Riemannian metric) for $X$ at the point $x(\theta )$.  In 
this way we can view $L_{C^0}X$ as a Riemannian manifold.  

In the second expression in $(1.1)$ for $\Cal A$, the first term is 
the usual kinetic energy for a path in the Riemannian 
manifold $L_{C^0}X$, and the second term represents a 
potential energy term, corresponding to the energy 
function on the finite energy loop space $L_{W^1}X$, 
$$E(x:S^1\to X)=\frac 12\int_{S^1}\langle dx\wedge *dx\rangle =\frac 
12\int\vert\frac {\partial x}{\partial\theta}\vert^2d\theta .\tag 1.3$$
Note that the Riemannian metric $(1.2)$ and $E$ depend upon 
the radius of $S^1$.  

From $(1.1)-(1.3)$ we can deduce, in a rough heuristic way, 
that the quantum Hamiltonian for the sigma model is of 
the form 
$$H=\Delta +E\tag 1.4$$
where $\Delta$ is the Laplacian for the Riemannian manifold 
$L_{C^0}X$, and $E$ is viewed as a (extremely singular) 
multiplication operator.  At this heuristic level, the 
operator $H$ should define a nonnegative self-adjoint 
operator on a Hilbert space that is of the form 
``$L^2(LX,dV)$'', where $dV$ is a fictional Riemannian volume 
element.  

To rigorously define $H$, one must introduce a cutoff 
which breaks scale covariance, and (as discussed in the 
introduction) it is expected that in general, after 
removing the cutoff, there is a residual nontrivial 
dependence of $H$ on $R$, the radius of the circle.  

One can at least schematically think of one aspect of 
this renormalization process in terms of a commutative 
diagram (involving unbounded operators) 
$$\matrix L^2(\Omega^2_RdV)&@>{\Omega\circ (\Delta +E)\circ\Omega^{
-1}}>>&L^2(\Omega^2_RdV)\\
\uparrow\Omega^{-1}&&\uparrow\Omega^{-1}\\
L^2(dV)&@>{\Delta +E}>>&L^2(dV)\endmatrix ,\tag 1.5$$
where $\Omega =\Omega_R$, the fictional ground state for $\Delta 
+E$ 
(depending upon the radius $R$ of the circle), is viewed as 
a multiplication operator.  In this diagram the coupled 
pair $\Omega^2_RdV$ should represent a well-defined finite positive 
measure (when topological terms are added to the action, 
this might more generally represent a measure having 
values in a line bundle).  The point is that to obtain a 
well-defined Hamiltonian, one must consider states 
relative to the ground state.  

\smallskip

\flushpar Example (see \S 1.3 of [Ga]).  Suppose $X=\Bbb R$, and 
for simplicity, we add an explicit mass term $m^2\vert x(t,\theta 
)\vert^2$ 
to the integrand in $(1.1)$.  By expanding $x(t,\theta )=\sum x_k
(t)e^{ik\theta}$ 
in a Fourier series, one sees that $\Delta +E$ can be written 
as a sum of oscillators.  The formal Hilbert space 
$L^2(L\Bbb R,dx)$ and the ground state 
$$\Omega =exp(-\frac 12\sum_{k=-\infty}^{\infty}\sqrt {m^2+k^2}\vert 
x_k\vert^2)\tag 1.6$$
do not make sense individually.  But in the top row of 
the diagram 
$$\matrix L^2(\frac 1{\Cal Z}e^{-\sum\sqrt {m^2+k^2}\vert x_k\vert^
2}dx)&@>{\Omega\circ (\Delta +E)\circ\Omega^{-1}}>>&L^2(\frac 1{\Cal Z}
e^{-\sum\sqrt {m^2+k^2}\vert x_k\vert^2}dx)\\
\uparrow\Omega^{-1}&&\uparrow\Omega^{-1}\\
L^2(L\Bbb R,dx)&@>{\Delta +E}>>&L^2(L\Bbb R,dx)\endmatrix \tag 1.7$$
the measure is a well-defined Gaussian measure, and the 
operator can be rigorously defined.  

\smallskip

Before discussing what is mathematically known about 
operators on $LX$ (when $X$ is curved), we digress to recall 
an idea inspired by Witten's work.  Suppose that $Y$ is a 
finite dimensional Riemannian manifold (but we will want 
to heuristically apply this to $Y=LX$), and that $\Cal E$ is a 
(Morse) function on $Y$.  There is a commutative diagram 
(involving unbounded operators) 
$$\matrix L^2(Y,e^{-\beta \Cal E}dV)&@>{\Delta_{\beta}}>>&L^2(Y,e^{
-\beta \Cal E}dV)\\
\uparrow e^{\frac {\beta}2\Cal E}&&\uparrow e^{\frac {\beta}2\Cal E}\\
L^2(Y,dV)&@>{\Delta_Y}>>&L^2(Y,dV)\endmatrix ,\tag 1.8$$
akin to $(1.5)$, where $\Delta_{\beta}=e^{\frac {\beta}2\Cal E}\circ
\Delta_Y\circ e^{-\frac {\beta}2\Cal E}$, and $dV$ denotes 
the Riemannian volume element (a refinement of this, 
involving the conjugation of $d$, the exterior derivative, 
is relevant to Morse theory and the supersymmetric 
sigma model; see [Wi1]).  

In the case $Y=L_{W^1}X$, there is a natural choice for $\Cal E$, 
$\Cal E=E$, there is an analogue of $e^{-\beta \Cal E}dV$, namely Wiener 
measure with inverse temperature $\beta$, denoted $\nu_{\beta}$ 
(although note this measure is not supported on $Y$), and 
there is an analogue of $\Delta_{\beta}$, which has been investigated 
by Gross and others.  From the sigma model point of 
view, one can view these objects as regularizations of 
the heuristic expressions that we introduced above.  
This motivates the study of possible limits of Wiener 
measure $\nu_{\beta}$, and $\Delta_{\beta}$, plus a potential function that might 
stand in for $E$, as $\beta\downarrow 0$ (but note that $\Delta_{
\beta}$ by itself tends 
to the Laplacian for the $W^1$ loop space, not the $C^0$ loop 
space).  Note also that letting $\beta\downarrow 0$ corresponds to $
R\uparrow\infty$.  

Now suppose that $X=K$, a compact simply connected Lie 
group with a simple Lie algebra, $\frak k$.  This space has an 
essentially unique biinvariant Riemannian structure, 
determined by a multiple of the Killing form, which we 
will $\underline {normalize}$ in a standard way (the length squared of 
a long root is $2$; in the case $K=SU(2,\Bbb C)$$=S^3$, this 
means the inner product is $\langle x,y\rangle =tr_{\Bbb C^2}(x^{
*}y)$, for 
$x,y\in su(2,\Bbb C)$).  Eliminating this normalization would 
introduce a mass parameter.  

A first fact of note is that Gross has proven that for a 
large class of potentials, $\{V\}$, $\Delta_{\beta}+V$ has a unique ground 
state (see [Gr]; the nature of the spectrum apparently 
remains unknown).  I am unaware of any results 
concerning limits of these operators as $\beta\downarrow 0$, which in the 
present context amounts to removing a regularization.  
Our speculations to follow are possibly related to these 
limits (and hence to an infinite radius limit).  

Let $G$ denote the complexification of $K$ (if $K=SU(2)$, 
then $G=SL(2,\Bbb C)$).  As I will describe in $\S 2$, there is a 
natural completion of the loop space $LG$, the 
hyperfunction loop space $L_{hyp}G$, with the properties 
that (1) $LK$ acts from the left and right, and (2) the 
Wiener measures $\nu_{\beta}$ converge to a biinvariant probability 
measure $\mu$ on $L_{hyp}G$ as $\beta\downarrow 0$.  This depends in an 
essential way on our assumption that $K$ is simply 
connected.  The measure $\mu$ should be characterized in the 
following way:  

\proclaim{(1.9) Conjecture}There is a $\underline {unique}$ probability 
measure on $L_{hyp}G$ which is biinvariant with respect to 
$LK$ (for uniqueness it should suffice to consider 
polynomial loops).  
\endproclaim\ 

In this paper we assume that the Hilbert space for the 
sigma model with target space $K$, in the infinite radius 
limit, is $L^2(L_{hyp}G,\mu )$, and we propose to use the 
structure of $L_{hyp}G$ and $\mu$ to infer properties of the 
Hamiltonian $H=H_{\infty}$.  

\smallskip

\flushpar$(1.10)$ Remarks.  (a) We emphasize that $\mu$ is a 
probability measure.  Even in a heuristic sense, $\mu$ is not 
to be confused with the fictional Riemannian volume 
element $dV$ on $L_{C^0}K$.  We think of switching from $dV$ to 
$\mu$ as similar to the vacuum renormalization process in 
$(1.5)$, with $R=\infty$.  

(b) In $\S 4$ we will discuss the $R<\infty$ case, as best we 
understand it.  At this time it is not clear how to 
formulate a characterization of the corresponding 
measure, $\Omega_R^2dV$ (assuming it exists), similar to $(1.9)$.  

\smallskip

A generic $g\in L_{hyp}G$ can be represented as a formal 
product 
$$g=g_{-}\cdot g_0\cdot g_{+},\tag 1.11$$
where $g_0\in G$ is constant, and $g_{\pm}$ are $G$-valued 
holomorphic functions on the disks $\Delta =\{\vert z\vert <1\}$ and 
$\Delta^{*}=\{\vert z\vert >1\}$, respectively, with $g_{+}(0)=1$ and $
g_{-}(\infty )=1$.  
If $g\in L_{C^0}G$ (an ordinary continuous loop in $G$) and 
generic (i.e.  the Toeplitz operator associated to $g$ is 
invertible), then $(1.11)$ is the standard triangular or 
Riemann-Hilbert or Birkhoff factorization of $g$ (see [CG] 
or chapter $8$ of [PS]).  

There is a strongly motivated conjecture for the $g_0$ 
distribution of $\mu$; in the case $K=SU(2)$, the conjecture 
states that 
$$(g_0)_{*}\mu =\frac 1{\Cal Z}tr(g_0^{*}g_0)^{-3}dm(g_0),\tag 1.12$$
where $dm$ denotes an invariant measure for $SL(2,\Bbb C)$, and 
$\Cal Z$ normalizes the total mass to be one (see $(3.4)$ below 
for the general formula).  
 
We now introduce two further assumptions.  The first 
is that the low energy states of the sigma model should 
be functions of $g_0$ alone.  The second is that the 
Hamiltonian, or an approximation to it, should act on the 
space $L^2(G,(g_0)_{*}\mu )$, and that this approximation should be 
given by a second order operator, necessarily biinvariant 
with respect to $K$.  We will refer to this approximation 
as the ``zero-mode Hamiltonian'', since functions of $g_0$ 
are rotation invariant (but note that $L^2(G,(g_0)_{*}\mu )$ is 
properly contained in the space of all rotation invariant 
functions!).  In the case of $K=SU(2)$, $K$ biinvariance 
leaves just one radial degree of freedom for the radial 
part, and as we will calculate in $ $\S 3, this leads to $(0.1)$.  

We will now briefly indicate how this generalizes to 
include other action terms.  

Returning to a general target space $X$, given a ``$B$-field'' 
on $X$ (i.e.  an element $b\in\hat {H}^2(X,\Bbb T)$, the degree $
2$ 
Cheeger-Simons differential characters, which can be 
written heuristically as $b=exp(2\pi iB)$, where $B$ is a 
$2$-form on $X$), there is a multivalued generalization of 
the sigma model action, which gives rise to well-defined 
Feynmann amplitudes, 
$$exp(-\beta \Cal A(x)+2\pi i\int_{\Sigma}x^{*}B).\tag 1.13$$
The deformation invariant of a $B$-field is the cohomology 
class of $dB$ in $H^3(X,\Bbb Z)$.  

In the case $X=K$, there are special $B$-fields, the WZW 
action terms, which are parameterized by a level 
$l\in \Bbb Z=H^3(K,\Bbb Z)$.  When the inverse temperature 
parameter $\beta$ and the level $l$ satisfy $\beta =l$, then the 
corresponding sigma model is the conformally invariant 
WZW model at level $l$, for which the Hilbert space is 
$$H^0_{L^2}(\Cal L^{*\otimes l}),\tag 1.14$$
the space of holomorphic sections of the line bundle 
$\Cal L^{*\otimes l}$, where $\Cal L=\hat {L}_{hyp}G\times_{\Bbb C^{
*}}\Bbb C$, and the vacuum is (an 
appropriate power of) the Toeplitz determinant, $detA(\hat {g})$, 
viewed as a section (it is worth noting that in this 
exceptional example, vacuum renormalization, as in $(1.5)$, 
is not necessary).  It is well-known that this space 
decomposes discretely into irreducible representations 
with respect the action of $\hat {L}K\times\hat {L}K$ (the Kac-Moody 
extensions of the loop groups), and one can use this to 
find the (discrete) spectrum of the model.  In this 
conformally invariant context the various assumptions 
we made above, about the dependence of low energy 
states on $g_0$ and so on, are, in some sense, known to be 
correct (see [Ga]).  

We consider the ansatz that the Hilbert space for the 
corresponding massive deformation at level $l$ is the 
larger space of all sections, 
$$\Omega^0_{L^2}(\Cal L^{*\otimes l}).\tag 1.15$$
How the orthogonal complement of the discrete part 
$(1.14)$ decomposes, if at all, is simply not known.  
However it is again reasonable to investigate the 
possibility that in terms of the Riemann-Hilbert 
factorization $(1.11)$, the low energy states depend only 
upon $g_0$, and so on.  We will write down the conjectural 
radial part of the zero-mode Hamiltonian at level $l$ in \S 3.  

\bigskip

\centerline{\S 2. The Structure of $\mu$.}

\bigskip

The existence of biinvariant limits of the measures $\nu_{\beta}$ 
as $\beta\downarrow 0$ is proven in [Pi].  Here I will give an outline of 
a relatively direct proof.  The argument is broken into 
seven steps, two of which are listed as conjectural.  
Conjectural step 5 can be dispensed with; the results of 
[Pi] can be used to bypass this step, but this detour is 
long and step 5 is of considerable intrinsic interest.  
Conjectural step 7, which gives an explicit formula for 
the $g_0$ distribution of $\mu$, is essential for the purposes of 
this paper.  

By definition (see chapter 2, Part III, of [Pi]), as a set, 
$$L_{hyp}G=G(\Cal O(S^{1-}))\times_{G(\Cal O(S^1))}G(\Cal O(S^{1+}
)),\tag 2.1$$
where $G(\Cal O(S^1))$ is the group of analytic loops in $G$, 
$G(\Cal O(S^{1+}))$ is the direct limit of the groups 
$G(\Cal O(\{r<\vert z\vert <1\})$ as $r\uparrow 1$ ($G$-valued holomorphic functions 
on some annulus just inside $S^1$), $G(\Cal O(S^{1-}))$ is the direct 
limit of the $G(\Cal O(\{1<\vert z\vert <r\})$ as $r\downarrow 1$, and $
G(\Cal O(S^1))$ acts on 
these latter two groups by multiplication.  This is a 
nonabelian generalization of Sato's realization of the dual 
of $\Cal O(S^1)$ (the elements of this dual are called 
hyperfunctions, and generalize the notion of a 
distribution).  From this global definition it is clear that 
$G(\Cal O(S^1))$ acts on the left and right of $L_{hyp}G$ (but this 
action is far from transitive).  The set $L_{hyp}G$ can be 
turned into a complex manifold, where a model 
coordinate neighborhood is given by $(1.11)$; the coordinates 
for this neighborhood are 
$$(\theta_{-},g_0,\theta_{+})\in H^1(\Delta^{*},\frak g)\times G\times 
H^1(\Delta ,\frak g)\tag 2.2$$
where $\theta_{+}=g_{+}^{-1}\partial g_{+}$ and $\theta_{-}=(\partial 
g_{-})g_{-}^{-1}$.  Other neighborhoods 
are obtained by translation by elements of $G(\Cal O(S^1))$.

\smallskip

\flushpar$(2.3)$Technical Remark.  Below it will 
occasionally be useful to replace $\theta_{+}$ by its integral 
$x_{+}\in H^0(\Delta ,\frak g)_0$, where $\theta_{+}=\partial x_{
+}$, $x_{+}(0)=0$.  One could 
imagine using other coordinates as well.  But $(2.2)$ is 
natural in the following sense:  there is a natural action 
of $Diff^{+}_{C^{\omega}}(S^1)$ on $L_{hyp}G$, in addition to the action of 
$LK\times LK$; the coordinates $(2.2)$ are equivariant with 
respect to the subgroup $PSU(1,1)$, where $PSU(1,1)$ acts 
naturally on $H^1(\Delta ,\frak g)$.  Assuming the truth of our 
conjecture $(1.9)$, the measure $\mu$ is invariant with respect 
to these actions.  

\smallskip

There is a natural inclusion of $L_{C^0}G\to L_{hyp}G$; this 
follows from the existence of Riemann-Hilbert 
factorization for continuous loops.  Wiener measure $\nu_{\beta}$ 
on $L_{C^0}K$ can therefore be viewed as a probability 
measure on $L_{hyp}G$.  We recall that $\nu_{\beta}$ is characterized 
in the following way:  given vertices $\{v\}$ and associated 
edges $\{e\}$ around $S^1$, the distribution of the values $\{g(v
)\}$ 
is given by the probability measure on $\prod_{\{v\}}K$:  
$$\frac 1{\Cal Z}\prod_{\{e\}}p_{Tl(e)}(g_{\partial e})\prod_{\{v
\}}dg_v,\tag 2.4$$
where $T=1/\beta$, $p_t$ denotes the heat kernel for $K$ [in 
particular $p_t(g,h)\sim\frac 1{\Cal Z}exp(-\frac 1{2t}d(g,h)^2)$ as $
d(g,h)\to 0$], and 
$g_{\partial e}$ denotes the pair of values of $g$ at the ends of the 
edge $e$.  Unfortunately this characterization is not 
directly useful in understanding $\nu_{\beta}$ in terms of the 
Riemann-Hilbert coordinates $\theta_{-},g_0,\theta_{+}$.  

We now turn to the basic steps of the argument.  

\proclaim{Step 1} $\nu_{\beta}$ is quasiinvariant with respect to 
$L_{W^1}K$ (finite energy loops) acting on $L_{C^0}K$ from either 
the left or 
right.\endproclaim

\proclaim{Step 2} $\nu_{\beta}$ is asymptotically invariant as $\beta
\downarrow 0$ 
in the following precise sense:  for each $p<\infty$, given 
$g'\in L_{W^1}K$, 
$$\int_{LK}\vert\frac {d\nu_{\beta}(g'g)}{d\nu_{\beta}(g)}-1\vert^
pd\nu_{\beta}(g)\le 2c(\beta )\Gamma (\frac {p+1}2)(2\beta E(g'))^{
p/2},$$
where $c(\beta )\to 1$ as $\beta\to 0$.  There is a similar estimate for 
$g'$ acting on the right.  
\endproclaim

\proclaim{Step 3}With $\nu_{\beta}$ probability one, $g$ has a 
Riemann-Hilbert factorization as in $(1.11)$, and $g_{\pm}$ and $
x_{\pm}$ 
have the same ``smoothness properties'' as $g$, where 
$\partial x_{+}=g_{+}^{-1}\partial g_{+}$, $x_{+}(0)=0$.  
\endproclaim

``Smoothness'' in Step 3 can be understood in various 
ways.  A version sufficient for our purposes is the 
following.  It is known that with $\nu_{\beta}$ probability one, $
g$ 
has a derivative of order $s$ in a Sobolev (or Holder) 
sense, for any $s<1/2$.  According to Step 3, the same is 
true for $g_{\pm}$ and $x_{\pm}$.  In particular we have 
$$\sum_{n>0}n^{\alpha}\vert\hat {x}_{+}(n)\vert^2<\infty ,\quad a
.e.\quad [\nu_{\beta}],\tag 2.5$$
for each $\alpha <1$.  

These first three steps are true for an arbitrary 
compact type Lie group $K$.  In particular the first two 
steps involve a reduction to a linear situation via the 
use of stochastic analysis (see \S 4.1 of Part II of [Pi]).  
The third step depends fundamentally on the fact that 
the conjugation operator is continuous on the class of 
Holder continuous functions, $C^{\mu}$, for any $0<\mu <1$ (see $
2.$ 
on p $60$ of [CG]).  

The next step depends crucially upon the simple 
connectedness of $K$.  In the case $K=SU(2)$ case, if we 
write 
$$g_{+}(z)=1+\left(\matrix a_1(g)&b_1(g)\\
c_1(g)&-a_1(g)\endmatrix \right)z+\hat {g}_{+}(2)z^2+...\tag 2.6$$
a straightforward calculation shows that for 
$k=\left(\matrix a&bz\\
-\bar {b}z^{-1}&\bar {a}\endmatrix \right)\in SU_2^{\tau}\subset 
LSU_2$, 
$$b_1(gk^{-1})=\frac {ab_1(g)-b}{\bar {b}b_1(g)+\bar {a}}.\tag 2.7$$
In other words, the right action of $k\in SU_2^{\tau}$ on $g\in L_{
hyp}G$ 
intertwines with the natural linear fractional action of 
$SU_2$ on $b_1\in\hat {\Bbb C}$ ($SU_2^{\tau}$ is a subgroup of $
LSU_2$ which is 
conjugate to $SU_2$ via an outer automorphism $\tau$, hence 
the notation).  This latter action is transitive and 
completely determines the form of an invariant measure.  
Since the $\nu_{\beta}$ are asymptotically invariant, the $b_1$ 
distributions are asymptotically invariant with respect 
to this $SU_2$ action on $\hat {\Bbb C}$.  This leads to the following 
conclusion.  

\proclaim{Step 4}In the limit as $\beta\to 0$, the distribution of 
$b_1$ is the $SU_2$-invariant distribution on $\hat {\Bbb C}$, 
$$\frac 1{\Cal Z}(1+\vert b_1\vert^2)^{-2}dm(b_1).$$
\endproclaim

The behavior of the measures $(b_1)_{*}\nu_{\beta}$ contrasts sharply 
with the behavior of the Gaussian measures $\frac 1{\Cal Z}exp(-\beta 
x^2)dx$ 
on Euclidean space (which is what we encounter for 
$K=T$, a flat torus), as $\beta\to 0$, because the ``probabilistic 
mass'' of the latter measures escapes to $\infty$ as $\beta\to 0$.  One 
theme of this note is that the preservation of 
probabilistic mass, which depends essentially on the 
semisimplicity of $K$, is related to the existence of a 
mass gap for the sigma model.  

\smallskip

\flushpar$(2.8)$ Remarks.  (a) For a general simply 
connected $K$, there is a result similar to step 4, where 
$b_1$ is replaced by the coordinate for the highest root 
space of $\frak g$.  

(b) Note that $\theta_0=\hat{\theta}_{+}(0)=\hat {x}_{+}(1)=\hat {
g}_{+}(1)\in \frak g$.  Conjecturally, 
$$\lim_{\beta\to 0}(\theta_0)_{*}\nu_{\beta}=\frac 1{\Cal Z}(1+\vert
\theta_0\vert^2)^{-d-1}dm(\theta_0),\tag 2.9$$
where $d=dimn_{\Bbb C}(\frak g)$.  But at this point there does not 
exist even a conjectural explicit formula for the joint 
distribution of all the modes $\theta_0$, $\theta_1=\hat{\theta}_{
+}(1)$,...  

\smallskip

Using invariance we can now use Steps 3 and 4 to show 
that the distributions of all the coefficients for $g_{+}$ (or 
$x_{+}$ or $\theta_{+}$) assume a finite shape as $\beta\to 0$.  This is proven 
in [Pi], using an induction argument.  I believe, however, 
that there is a more elegant explanation, possibly useful 
in a more general context.  A corollary of $(2.4)$ is that 
for fixed $\alpha <1$, and for each $R>0$, 
$$\nu_{\beta}\{n^{\alpha}\vert\hat {x}_{+}(n)\vert^2>R\}\to 0\quad 
as\quad n\to\infty .\tag 2.10$$
It is natural to ask whether there exists $\alpha$ such that 
the sequence in $(2.10)$ is actually nonincreasing (but not 
necessarily going to $0$), for all $\beta$ and $R$; if so there 
exists a largest such $\alpha$, $\alpha_c$.  In the abelian case one can 
easily calculate that $\alpha_c=2$.  

\proclaim{Conjectural Step 5}For $\alpha =0$, $(2.5)$ is a 
nonincreasing function of $n$, for all $\beta >0$ and $R\ge 0$, i.e.  
$\alpha_c\ge 0$.  In particular 
$$\nu_{\beta}\{\vert\hat {x}_{\pm}(n)\vert >R\}\le\nu_{\beta}\{\vert
\hat {x}_{\pm}(1)\vert >R\}.$$
\endproclaim

The following is a consequence of Step 4 and 
(conjectural) Step 5.  

\proclaim{Step 6}There exists a constant $d$ (depending 
only upon $\frak g$) such that 
$$\lim_{\beta\to 0}\nu_{\beta}\{\vert\hat {x}_{\pm}(n)\vert >R\}\le\frac 
d{(1+(R/d)^2)}.\tag 2.12$$
\endproclaim

This step implies that the mass of the $\nu_{\beta}$ does not 
escape to $\infty$ as $\beta\to 0$, at least when we consider the $
\theta_{\pm}$ 
coordinates.  This has already been done in [Pi] in a 
qualitative way; the point of $(2.12)$ is to quantity this 
result in an elegant way.  To complete our outline, we 
need to know that mass does not escape to infinity 
through $g_0$.  Again, this has already been done in a 
qualitative way in [Pi], but we need an explicit formula.  

Suppose that we choose a maximal torus $T$ for $K$, and a 
choice of positive roots for the action of the 
corresponding Cartan subalgebra $\frak h$ of $\frak g$.  We can 
generically write $g_0\in G$ in triangular form, $g_0=l_0m\bold a
u_0$, 
where we have further decomposed the diagonal term 
into a phase $m\in T$ and its magnitude $\bold a\in exp(\frak h_{
\Bbb R})$.  

\proclaim{Conjectural Step 7} We have 
$$\lim_{\beta\downarrow 0}\int \bold a(g)^{-i\lambda}d\nu_{\beta}^{
*=*}(g)=\prod_{\alpha >0}\frac {sin(\frac {\pi}{2\dot {g}}\langle
\rho ,\alpha\rangle )}{sin(\frac {\pi}{2\dot {g}}\langle\rho -i\lambda 
,\alpha\rangle )}\tag 2.13$$
where $\dot {g}$ is the dual Coxeter number, $\rho$ is the sum of the 
positive roots, and $\lambda\in \frak h_{\Bbb R}^{*}$ (and recall that the inner 
product has been normalized).  
\endproclaim

In the case of $K=SU_2$, this is equivalent to $(1.12)$.  The 
original motivation for this conjecture is explained in 
$\S 4.4$ of Part III of [Pi].  This formula should be 
compared with the known formula of Harish-Chandra, 
$$\lim_{\beta\uparrow\infty}\int \bold a(g)^{-i\lambda}d\nu_{\beta}
(g)=\bold c(\rho -i\lambda )=\prod_{\alpha >0}\frac {\langle\rho 
,\alpha\rangle}{\langle\rho -i\lambda ,\alpha\rangle}\tag 2.14$$
(see $\S 4.4$ of Part II of [Pi]).  

When we incorporate the level $l$, the generalization of 
conjectural step $8$ is 
$$\lim_{\beta\downarrow 0}\int \bold a^{-i\lambda}d\nu_{\beta ,l}
=\prod_{\alpha >0}\frac {sin(\frac {\pi}{2(\dot {g}+l)}\langle\rho 
,\alpha\rangle )}{sin(\frac {\pi}{2(\dot {g}+l)}\langle\rho -i\lambda 
,\alpha\rangle )}\tag 2.15$$
As $l\to\infty$, we recover the classical limit of Haar measure, 
$(2.14)$.  If we write $(g_0)_{*}\mu_l=\phi_ldm(g_0)$, then $(2.1
5)$ is 
equivalent to the following formula for the 
Harish-Chandra transform:  
$$(\Cal H\phi_l)(\lambda )=c\prod_{\alpha >0}\frac {\langle -i\lambda 
,\alpha\rangle}{sin(\frac {\pi}{2(\dot {g}+l)}\langle -i\lambda ,
\alpha\rangle )}=\prod_{\alpha}\Gamma (1+i\frac {\pi\langle\lambda 
,\alpha\rangle}{2(\dot {g}+l)})\left(2.16\right)$$
(this follows from $(4.4.27)$ of Part II of [Pi]).  

Steps 6 and 7 imply that the measures $\nu_{\beta}$ have limits in 
$L_{hyp}G$ as $\beta\to 0$.  Asymptotic invariance implies that 
these limits are biinvariant with respect to analytic 
loops in $K$.  The remaining step is to show that there is 
a unique such measure.  Considerable progress has been 
made, but this question remains open.  

\smallskip

\flushpar$(2.17)$Remark.  Although not directly relevant in 
this paper, we mention that there are conjectural 
expressions for the $\theta_{\pm}$ distributions, at least in terms of 
other, more explicit, limits.  For example, for $K=SU(n)$ 
in the defining representation, conjecturally 
$$(\theta_{-})_{*}\mu_l=\lim_{n\to\infty}\frac 1{\Cal Z}det(1+Z^{
*}Z)^{-2-l}dm(P_n\theta_{-}),\tag 2.18$$
where $Z=Z(g_{-})=C(g_{-})A(g_{-})^{-1}$ (following the notation in 
[PS]), $g_{-}$ corresponds to $P_n\theta_{-}$, and $P_n$ projects $
\theta_{-}$ to its 
first $n$ coefficients (so that it is an orthogonal 
projection for $H^1(\Delta^{*},\frak g)$).  This expression is manifestly 
$PSU(1,1)$ invariant. This is the analogue of a well-known 
formula of Harish-Chandra for the invariant measure on 
a finite dimensional flag space (see [Helg], Thm 5.20, p 
198).

\bigskip

\centerline{\S 3. The Conjecture for the Radial Part ($R=\infty$).}

\bigskip

We introduce the ansatz that the subspace 
$$L^2(G,(g_0)_{*}\mu )\subset L^2(\mu )\tag 3.1$$
is invariant, or at least approximately invariant, with 
respect to the action of the Hamiltonian, $H=H_{\infty}$.  This 
approximation, $H_G$, will necessarily be a $K\times K$-invariant 
linear operator.  To further restrict the possibilities, we 
also assume that $H_G$ is a second order differential 
operator.  

Consider the Cartan decomposition 
$$\psi :K\times \frak p\to G:k,x\to g=ke^x.\tag 3.2$$
In these coordinates Harish-Chandra's formula for the 
Haar measure of $G$ is 
$$dg=\prod_{\alpha >0}\vert\frac {sinh\alpha (\frak a(x))}{\alpha 
(\frak a(x))}\vert^2dk\times dx\tag 3.3$$
where $x\in \frak p$ is $K$-conjugate to $\frak a(x)\in \frak h_{
\Bbb R}$, and the product 
is over the positive roots (see [Helg], Thm 5.8, p 186).  

Conjectural Step $7$ of $\S 2$ is equivalent to 
$$(g_0)_{*}\mu =\frac 1{\Cal Z}\frac {\sum_W(-1)^w\int\prod_{\alpha 
>0}\langle\lambda ,\alpha\rangle^2sinh(\pi\langle\lambda ,\alpha\rangle 
/2\dot {g})^{-1}a^{iw\cdot\lambda}d\lambda}{\prod_{\alpha >0}(a^{
\alpha}-a^{-\alpha})}dg_0.\tag 3.4$$
where in this formula, for $g\in G$, $KgK=Ka(g)K$, 
$a\in exp(\frak h_{\Bbb R})/W$.  This reduces to $(1.12)$ for $K=
SU_2$.  

If $K=SU(2,\Bbb C)$, then $\frak p$ consists of $2\times 2$ Hermitian 
matrices, and we can use the standard identification 
$$K\times \frak p=S^3\times (\Bbb R\vec{\imath }+\Bbb R\vec{\jmath }
+\Bbb R\vec {k}),\tag 3.5$$
where $\vec{\imath}\leftrightarrow\left(\matrix 1&0\\
0&-1\endmatrix \right)$, $\vec{\jmath}\leftrightarrow\left(\matrix 
0&1\\
1&0\endmatrix \right)$, and $\vec {k}\leftrightarrow\left(\matrix 
0&i\\
-i&0\endmatrix \right)$.  In these 
coordinates 
$$dg=(\frac {sinh(2\vert x\vert )}{2\vert x\vert})^2dk\times dx,\tag 3.6$$
where $dk$ denotes Haar measure for $SU(2)$ and $dx$ is 
Lebesgue measure for $\Bbb R^3$.  The formula $(1.12)$ is then 
$$(g_0)_{*}\mu_0=\frac 1{\Cal Z}\frac 1{cosh^3(2\vert x\vert )}(\frac {
sinh(2\vert x\vert )}{2\vert x\vert})^2dk\times dx$$
$$=\frac 1{\Cal Z}sech(2\vert x\vert )(\frac {tanh(2\vert x\vert
)}{2\vert x\vert})^2dk\times dx$$
$$=\frac 1{\Cal Z}\delta (r)dr\times dk\times dA_{S^2}(x'),\tag 3.7$$
where $\delta =sech(r)tanh^2(r)$, $2x=rx'$, $x'\in S^2$ (see \S 4.4 of 
Part III of [Pi]).  

Let $D=\frac {\partial}{\partial r}$, and let $H_r$ denote the radial part of $
H_G$.  
Since $H_r$ is self-adjoint and nonnegative with respect to 
$\delta (r)dr$, and because $H$, hence $H_r$, applied to a constant 
(the vacuum) vanishes, $H_r$ must necessarily be of the 
form 
$$H^{\alpha}_r=-\delta^{-1/2}\circ D\circ\alpha (r)\circ D\circ\delta^{
1/2}+\frac {D(\alpha D\delta^{1/2})}{\delta^{1/2}}$$
$$=\alpha [-\delta^{-1/2}\circ D^2\circ\delta^{1/2}+\frac {D^2(\delta^{
1/2})}{\delta^{1/2}}]-D(\alpha )D$$
$$=\alpha [-\delta^{-1/2}\circ D^2\circ\delta^{1/2}+\frac 14-\frac {
15}4sech^2(r)]-D(\alpha )D\tag 3.8$$
where $\alpha =\alpha (r)$ is a positive function.  This is our 
initial ballpark conjecture.  

The principal symbol of $H_r^{\alpha}$, in the coordinate $r$, is $
\alpha\xi^2$, 
where $\xi$ is a variable dual to $r$.  Determining $\alpha$ is thus 
equivalent to picking out a preferred geometry.  We will 
now explain why $\alpha =1$ appears to be a preferred choice.  

We are assuming that $H_r$ is the radial part of an 
operator $H_G$, and there is an intermediate operator $H_{\frak p}$, 
acting on functions of $\frak p$ alone, in the Cartan 
decomposition (these are functions which are invariant 
with respect to the left action of $K$; we could just as 
well consider the right action).  The principal symbol of 
$H_{\frak p}$ corresponds to a metric on $\frak p$.  

In considering interesting possibilities for the principal 
symbol of $H_{\frak p}$, it seems that this metric has the form 
$$g_x(v,w)=\langle A(ad(x))v,w\rangle ,\tag 3.9$$
where $A$ is an analytic function which is expressible as 
a power series in powers of $ad(x)$, $x\in \frak p$.  In the 
appendix to this section, we will show that in all such 
cases, $\alpha =1$ (see (f) of Lemma $(A.2)$).  

Suppose that $\alpha =1$.  In this case $H_r$ is equivalent to 
$$-D^2+\frac 14-\frac {15}4sech^2(r),\tag 3.10$$
acting on odd functions of $r$.  The restriction to odd 
functions of $r$ is necessitated by the fact that 
$\delta^{1/2}=sech^{1/2}(r)tanh(r)$ is an odd function (functions in 
the domain of $H_r$ will then be of the form (odd function 
$/\delta^{1/2}$, which will represent a well-defined function on 
$G$).  This operator has a unique eigenvalue $\lambda =0$ 
corresponding to the ground state, $\delta^{1/2}$, and the rest of 
the spectrum is continuous and of the form $[m,\infty )$, 
where $m=\frac 14$ is the mass gap (Note:  if we remove the 
restriction on the domain of $(3.10)$ to odd functions, then 
the operator has a lower energy state, the even function 
$sech^{3/2}(r)$, which corresponds to the eigenvalue $-2$).  The 
scattering theory for the $sech^2$ potential (at least 
without domain restriction) is well-known (see e.g.  $\S 2.5$ 
of [L]).  Taking the domain restriction into account, this 
should be related to Zamolodchikov's conjectural 
$S$-matrix for this model (see [Z]).  

In our argument for $\alpha =1$, we noted that $H_r$ does not 
determine the form of $H_G$ (or $H_{\frak p}$).  At the level of $
G$, 
we have 
$$H_G=\Phi^{-1/2}\circ\Delta\circ\Phi^{1/2}-\frac {\Delta (\Phi^{
1/2})}{\Phi^{1/2}},\tag 3.11$$
where $\Delta$ (a Laplace type operator) is self-adjoint with 
respect to $dk\times dx$ and $(g_0)_{*}\mu =\Phi (dk\times dx)$.  There are 
numerous possibilities for $\Delta$.  

For example, relative to the Cartan decomposition 
$G=K\times \frak p$, we could have $\Delta =\Delta^K+\Delta^{\frak p}$, the sum of the 
Laplacians.  For this example the $m=1/4$ is directly 
related to the curvature of $G/K=H^3$ (relative to the 
normalization of our metric), because $\Delta^{G/K}$ is equivalent 
to $\Delta^{\frak p}+1/4$ (see Proposition 3.10 of [Helg], pg 268, and 
$(A.6)$ of the Appendix).  This is relevant to the 
explanation for various miracles that occur in harmonic 
analysis in $3$ versus $n$ dimensions (see [Helg], especially 
p 266).  

In the Appendix we consider a second possibility in 
detail.  This second possibility is interesting because it 
generalizes to other Riemannian manifolds, in a way 
which seems linked to renormalization of sigma models.  

We now discuss how to incorporate a level $l$, which 
presumably is related to the massive deformation of the 
conformally invariant WZW model at level $l>0$.  

We first recall from [Pi] that, at least conjecturally, 
$$(g_0)_{*}\mu_l=\frac 1{\Cal Z}\chi_{l/2}(e^{2x})\frac 1{cosh^3(
(2+l)\vert x\vert )}(\frac {sinh(2\vert x\vert )}{2\vert x\vert})^
2dk\times dx,\tag 3.12$$
where $\chi_{l/2}$ is a character, at least for integer $l/2$ [$\mu_
l$ 
denotes the measure gotten by coupling a certain density 
appropriate at level $l$; it is the conjectural limit of the 
$\nu_{\beta ,l}$ in $(2.15)$; see [Pi]).  

\smallskip

\flushpar[``Proof of $(3.12)$''.  In \S 4.4 of Part III of [Pi] we 
conjectured that (in the case $G=SL(2,\Bbb C)$, $\dot{\Lambda }=0$, and $
\lambda$ 
is identified with $\lambda\alpha_1$, $\alpha_1=\lambda_1-\lambda_
2$) 
$$\int_{L_{hyp}G}\bold a(g)^{-i\lambda}d\mu_l=\frac {sin(\frac {\pi}{
2(2+l)}2)}{sin(\frac {\pi}{2(2+l)}2(1-i\lambda ))}\tag 3.13$$
Write $(g_0)_{*}\mu_l=\phi_ldm(g_0)$, where $dm$ denotes $G$ Haar 
measure.  By (4.4.12) of [Pi] 
$$\Cal H\phi_l(\lambda )=\frac {i\lambda sin(\frac {\pi}{2+l})}{s
in(\frac {\pi}{2+l}i\lambda )}=sin(\frac {\pi}{2+l})\frac {\lambda}{
sinh(\frac {\pi}{2+l}\lambda )}\tag 3.14$$
By (4.4.15) of [Pi] 
$$\phi_l(\left(\matrix a&0\\
0&a^{-1}\endmatrix \right))=\frac 1{\Cal Z}\frac 1{(a^{2+l}+a^{-(
2+l)})^3}\frac {a^{2+l}-a^{-(2+l)}}{a^2-a^{-2}}$$
$$=\frac 1{\Cal Z}cosh^{-3}((2+l)x)\chi_{l/2}(\left(\matrix a^2&0\\
0&a^{-2}\endmatrix \right)),\tag 3.15$$
where $\chi_{l/2}$ is the character for the $SU(2)$ 
representation of dimension $l/2$ (assuming this is 
integral).  This implies $(3.12)$.]  

\smallskip

Write $r=(2+l)\vert x\vert$, $a=\frac 2{2+l}$, $D=\frac {\partial}{
\partial r}$, and 
$$\delta =sech^3(r)sinh(r)sinh(ar),\tag 3.16$$
so that the radial projection of $(g_0)_{*}\mu_l$ is (conjecturally) 
$\Cal Z^{-1}\delta (r)dr$.  

Assuming that $\alpha =1$, for the same reasons cited above, 
we find that $H^l_r$ is conjecturally of the form 
$$\delta^{-1/2}\circ\vert D\vert^2\circ\delta^{1/2}+\frac 14(5+2a^
2-6a\frac {tanh(r)}{tanh(ar)}-(acoth(ar)-coth(r))^2-15sech^2(r)).$$
The potential well for this operator digs deeper as $l\to\infty$, 
suggesting that the number of eigenvalues and bound 
states goes to $\infty$ as $l\uparrow\infty$.  

\bigskip

\bigskip

\centerline{\S 4. The Finite $R$ Case.}

\bigskip

As we explained in the introduction, and in $(1.5)$, for the 
sigma model with target $K$, we expect that there should 
be a natural Hilbert space $\Cal H(S^1_R)=L^2(\frac 1{\Cal Z}\Omega^
2_RdV)$ for each 
$0<R\le\infty$, where heuristically we think of $\Omega_R$ as the 
vacuum state.  

At this point we lack a construction, and a conjectural 
characterization (as in $(1.9)$), for the appropriate 
measure, when $R$ is finite.  However in this section we 
will assume this can be done.  The point of this section 
is to explore what appears to be a natural conjecture 
for the $g_0$ distribution.  

As in \S 3 we will write $g_0=l_0m\bold au_0$ for the triangular 
decomposition (when it exists), and $Kg_0K=KaK$ for the 
Cartan decomposition, where $\bold a\in A=exp(\frak h_{\Bbb R})$ and 
$a\in A/W$, respectively.  

As in Chapter XXI of [WW], $\theta_1$ will denote the odd theta 
function 
$$\theta_1(x,\tau )=2q^{1/8}sin(x)-2q^{3^2/8}sin(3x))+..\tag 4.1$$
$$=2q^{1/8}sin(x)\prod_{n=1}^{\infty}(1-q^n)(1-q^ne^{i2x})(1-q^ne^{
-i2x}),\tag 4.2$$
where $q=exp(2\pi i\tau )$ (this is the square of ``$q$'' in [WW]), 
$Im(\tau )>0$, and the equality is known as the Jacobi triple 
product formula.  This theta function has the 
quasi-periodicity properties 
$$\theta_1(x+\pi )=-\theta_1(x),\quad\theta_1(x+\tau )=-q^{1/2}e^{
-2ix}\theta_1(x),\tag 4.3$$
and zeroes at the points 
$$x=n\pi +m\pi\tau ,\quad m,n\in \Bbb Z.\tag 4.4$$
Below we will also need to consider the even theta 
functions $\theta_3$ and $\theta_4$, which have analogous properties 
(see [WW]).  

\proclaim{(4.5)Conjecture}The analogue of $(2.15)$ (the 
diagonal distribution) is 
$$\int \bold a^{-i\lambda}\frac 1{\Cal Z}\Omega^2_{R,l}dV=c\prod_{
\alpha >0}\{\frac {sinh(\frac {\pi}{2R(\dot {g}+l)}\langle\rho -i
\lambda ,\alpha\rangle )}{\langle\rho -i\lambda ,\alpha\rangle\theta_
1(\frac {\pi}{2(\dot {g}+l)}\langle\rho -i\lambda ,\alpha\rangle 
,iR)}\}\tag 4.6$$
where $c$ is determined by the condition that the right 
hand side of $(4.6)$ is $1$ at $\lambda =0$.  
\endproclaim

If we write $(g_0)_{*}(\frac 1{\Cal Z}\Omega_{R,l}^2dV)=\phi_{R,l}
(g_0)dm(g_0)$, where $dm$ 
denotes Haar measure for $G$, then $(4.5)$ is equivalent to 
$$(\Cal H\phi_{R,l})(\lambda )=c\prod_{\alpha >0}\{\frac {sin(\frac {
\pi}{2R(\dot {g}+l)}\langle\lambda ,\alpha\rangle )}{\theta_1(\frac {
\pi}{2(\dot {g}+l)}\langle i\lambda ,\alpha\rangle ,iR)}\}\tag 4.7$$
for the Harish-Chandra transform.  Note that the zeros 
of the sine function in $(4.7)$ exactly cancel with the 
zeros of $\theta_1(i(\cdot ))$, so the $\alpha$ factor in $(4.7)$ is smooth and 
rapidly decreasing as a function of the single variable 
$\langle\lambda ,\alpha\rangle$, for each positive root $\alpha$.  

The motivations for this conjecture are rather vague:  
the philosophy that theta functions are natural 
$q$-deformations of trigonometric functions, the relevance 
of the $q$-deformation of the affine algebra $\hat {L}\frak g$ to 
integrable models (see [Sm] and references there), and 
the surprising appearance of ``$\tau$'' in similar models 
(especially gauge theories; see e.g.  [Wi2]).  

To show that this formula is reasonable, there are 
several things that need to be checked.  The first is to 
note that $(4.6)$ reduces to $(2.15)$ when $R\uparrow\infty$.  This 
follows in an elementary way from $(4.1)$ (in verifying 
this, one must bear in mind the dependence of $c$ on $R$).  
Thus this formula is consistent with our earlier claim.  

Secondly we need to know that the transforms we are 
writing down actually correspond to positive measures.  
We first consider $(4.6)$.  

\proclaim{(4.8)Proposition}The right hand side of $(4.6)$ is 
a positive definite function of $\lambda\in \frak h_{\Bbb R}^{*}$.  
\endproclaim

\flushpar Proof of $(4.8)$.  Products of positive definite 
functions are positive definite.  In $(4.6)$ we have a 
product over roots, and it suffices to show that each 
factor is positive definite as a function of one variable 
(the distance from $ker(\langle\cdot ,\alpha\rangle )$).  

The dual Coxeter number is given by $\dot {g}=1+\langle\rho ,\theta
\rangle /2$, 
where $\theta$ is the highest root and (we recall that) $\rho$ is the 
sum of the positive roots.  Using this and the fact that 
$\langle\rho ,\alpha\rangle\le\langle\rho ,\theta\rangle$, for each root $
\alpha$, we can write 
$$\frac {\pi}{2(\dot {g}+l)}(\langle\rho ,\alpha\rangle -i\langle
\lambda ,\alpha\rangle )=x_0+iy,\tag 4.9$$
where $0<x_0<\pi$, and $y$ is a scaling of the variable 
$\langle\lambda ,\alpha\rangle$.  Since scaling a positive definite function does not 
change its positivity, it therefore suffices to prove that 
$$\frac {sinh((x_0+iy)/R)}{\theta_1(x_0+iy,iR)(x_0+iy)/R}\tag 4.10$$
is a positive definite function of $y$.  

This is a consequence of the following striking result, 
which probably is known.  

\proclaim{ (4.11)Lemma} For $0<x_0<\pi$, 
$$\frac 1{2\pi}\int\frac 1{\theta_1(x_0+iy,iR)}e^{ipy}dy=\frac 1{
\theta_1'(0,iR)}\frac {\theta_4(\pi Rp/2,iR)}{e^{x_0p}+e^{(x_0-\pi 
)p}}.\tag 4.12$$
\endproclaim

\flushpar Proof of $(4.11)$.  This is a straightforward 
residue calculation.  Suppose that $p>0$.  The residues 
for the integrand on the LHS of $(4.12)$, as a function of 
complex $y$, occur at the points $y=m\pi R+i(x_0+n\pi )$, 
$n,m\in \Bbb Z$, $n\ge 0$.  Thus the LHS of $(4.12)$ equals 
$$i\sum_{m\in \Bbb Z,n\ge 0}\frac {exp(ip(m\pi R+i(x_0+n\pi )))}{
i\theta_1'(-n\pi +im\pi R)}\tag 4.13$$
Using the quasi-periodicity properties of $\theta_1$, $(4.3)$, we 
obtain 
$$\theta_1'(-n\pi +im\pi R)=(-1)^{n+m}q^{-m^2/2}\theta'_1(0).\tag 4.14$$
Thus $(4.13)$ equals 
$$\frac {e^{-x_0p}}{\theta_1'(0)}(\sum_{n=0}^{\infty}(-1)^ne^{-\pi 
pn})(\sum_{m\in \Bbb Z}(-1)^mq^{m^2/2}e^{iRpm}).\tag 4.15$$
The second sum is expressible in terms of $\theta_4$, and this 
implies $(4.11)$.//  

\smallskip

The Fourier transform of $sin(y)/y$ is essentially a 
characteristic function.  This together with the Lemma 
implies that the inverse Fourier transform of $(4.10)$, as 
a function of $p$, is the convolution of measures 
$$\frac 1{\theta_1'(0,iR)}\frac 1{e^{x_0p}+e^{(x_0-\pi )p}}\theta_
4(\pi Rp/2,iR)*e^{-x_0p}\chi_{[-1/R,1/R]}(p)/2R\tag 4.16$$

The crucial fact now is that the function $\theta_4(x,iR)$ is 
positive for $x\in \Bbb R$.  Thus both measures are positive, 
implying that the convolution is positive.  This 
completes the proof of $(4.8)$.//  

\smallskip

\flushpar(4.17)Remark.  Another possible approach to 
$(4.11)$ is to consider the Jacobi triple product formula 
for $\theta_1$, $(4.2)$, which corresponds to an (infinite) 
convolution product formula for $(4.11)$.  If we compute 
the inverse Fourier transform for the $n$th term, we find 
that for $0<x_0<\pi$, 
$$\frac 1{2\pi}\int\frac 1{1-2cos(2(x_0+iy))q^n+q^{2n}}e^{ipy}dy\tag 4.18$$
$$=\frac {sin(\pi nRp)}{sinh(\pi nR)(e^{x_0p}+e^{(x_0-\pi )p})}.\tag 4.19$$
which is highly oscillatory.  From this point of view the 
positivity of $(4.11)$ is surprising.  

\smallskip

We now consider the formula $(4.7)$ for the 
Harish-Chandra transform.  The abstract inversion 
formula is 
$$\phi_{R,l}(g_0)=\frac 1{\Cal Z\prod_{\alpha >0}(a^{\alpha}-a^{-
\alpha})}\sum_W(-1)^w\int\prod_{\alpha >0}\{\langle\lambda ,\alpha
\rangle^2\frac {sin(\frac {\pi}{2R(\dot {g}+l)}\langle\lambda ,\alpha
\rangle )}{\theta_1(\frac {\pi}{2(\dot {g}+l)}\langle\lambda ,\alpha
\rangle )}\}a^{iw\cdot\lambda}d\lambda\tag 4.20$$
One can change variables in the integrals to reduce the 
calculations to the case $l=0$.  

We will analyze this in the case $K=SU(2)$.  

\proclaim{ (4.21)Lemma} 
$$\frac i{2\pi}\int\frac y{\theta_1(iy)}\frac {sinh(iy/R)}{iy/R}e^{
ipy}dy=\frac {Rsinh(\pi /R)\theta_3(\pi Rp/2,iR)}{4\theta_1'(0)(s
inh^2(\pi /(2R))+cosh^2(\pi p/2))}.\tag 4.22$$
\endproclaim\ 

\flushpar Proof of (4.21).  This is another straightforward 
residue calculation.  Suppose that $p>0$.  As a function 
of the complex variable $y$, the singularities of the 
integrand on the LHS of $(4.22)$ occur at the points 
$y=m\pi R+in\pi$, $m,n\in \Bbb Z$, $n>0$.  Using the formula $(4.
14)$ 
to calculate the residues, we see that the LHS of $(4.22)$ 
equals 
$$=R\sum_{n>0,m}\frac {exp(-\pi pn+i\pi Rpm)sinh((-n\pi +imR\pi )
/R)}{(-1)^n(-1)^mq^{-m^2/2}\theta_1'(0)}\tag 4.23$$
$$=\frac R{\theta_1'(0)}(\sum_mq^{m^2/2}e^{ipRm})(\sum_{n>0}(-1)^
nsinh(-n\pi /R)e^{-n\pi p})\tag 4.24$$
$$=\frac R{2\theta_1'(0)}\theta_3(\pi Rp/2)\{\frac {e^{-\pi /R}}{
e^{\pi p}+e^{-\pi /R}}-\frac {e^{\pi /R}}{e^{\pi p}+e^{\pi /R}}\}\tag 4.25$$
After some elementary manipulations, this leads to 
$(4.22)$.//  

\smallskip

For the $SU(2,\Bbb C)$ case we employ the same notation as in 
$\S 3$.  Thus we identify $a\in exp(\frak h_{\Bbb R})$ with $\left
(\matrix e^x&\\
&e^{-x}\endmatrix \right)$, $x\in \Bbb R$, 
we identify $\lambda\in \frak h_{\Bbb R}$ with $\lambda\alpha_1$, where $
\lambda\in \Bbb R$ and $\alpha_1$ is the 
positive root for $sl(2,\Bbb C)$, and we write $r=2\vert x\vert$.  We 
then have 
$$\phi_{R,l}(g)=\frac 1{\Cal Z}\frac 1{sinh(2\vert x\vert )}(-\frac {
\partial}{\partial z}\{\frac {\theta_3(Rz,iR)}{sinh(\pi /(2R))^2+
cosh(z)^2}\}\vert_{z=(2+l)x/\pi}).\tag 4.26$$
where $\Cal Z$ is a normalization constant, so that the integral 
with respect to Haar measure for $SL(2,\Bbb C)$ is one.  

\proclaim{(4.27)Proposition}We have $\phi_{R,l}\medspace \ge 0$.  
\endproclaim

\flushpar Proof of $(4.27)$.  By doing the differentiation in 
$(4.26)$, we see that $(4.27)$ is equivalent to 
$$\frac {\partial}{\partial z}ln(\theta_3(Rz,iR)\le 2\frac {sinh(
z)cosh(z)}{sinh(\pi /(2R))^2+cosh(z)^2},\quad z>0.\tag 4.28$$
The LHS of $(4.28)$ has period $\pi /R$.  The function 
$\theta_3(Rz,iR)$ is decreasing on $[0,\pi /(2R)]$, so the LHS of 
$(4.28)$ is negative on this interval, and hence the claim 
is trivially true on this interval.  It is straightforward 
to check that the RHS of $(4.28)$ is an increasing function 
of $z$.  Thus it suffices to prove $(4.28)$ on the finite 
interval $[\pi /(2R),\pi /R]$ (Note this means that for values 
of $R$ on the order of $1$, one can with confidence simply 
look at the graph of $\theta_3(Rz,iR)/(sh(\pi /(2R))^2+ch(z)^2)$, and 
check that it is decreasing on the appropriate interval).  

The LHS of $(4.28)$ equals 
$$-4Rsin(2Rz)\sum_{n=1}^{\infty}\frac {q^{n-1/2}}{1+2cos(2Rz)q^n+
q^{2n}}.\tag 4.29$$
The maximum of this function of $\pi /(2R)\le z\le\pi /R$ is the 
same as the maximum of the function 
$$2Rsin(\theta )e^{\pi R}\sum_{n=1}^{\infty}\frac 1{cosh(2\pi Rn)
-cos(\theta )},\quad 0\le\theta\le\pi ,\tag 4.30$$
where $z=(2\pi -\theta )/(2R)$.  

We first derive an easy bound for $(4.30)$, which is 
sufficient for $R$ sufficiently large.  On the domain 
$0\le\theta\le\pi$, the function $sin(\theta )/(cosh(2\pi Rn)-cos
(\theta ))$ has a 
maximum value of $1/sinh(2\pi Rn)$, which is achieved at the 
point $\theta =\theta_{R,n}$ satisfying $cos(\theta )=cosh(2\pi R
n)^{-1}$.  Thus 
$(4.30)$ is bounded by 
$$2Re^{\pi R}\sum_{n=1}^{\infty}\frac 1{sinh(2\pi Rn)},\tag 4.31$$
which is a decreasing function of $R$.  It is easy to check 
that this is dominated by the minimum of the RHS of 
$(4.28)$, 
$$\frac {2sinh(\pi /(2R))cosh(\pi /(2R))}{sinh(\pi /(2R))^2+cosh(
\pi /(2R))^2},\tag 4.32$$
for $R$ sufficiently large (in fact for $R>1/20$ (using 
Maple, for example).  But $(4.31)$ diverges as $R\downarrow 0$, and so 
this does not work in general.  

Now consider small $R$.  The function $(4.30)$ vanishes at 
$0$ and $\pi$, and it has a unique maximum at a point $\theta_R$ in 
the interior.  This point is determined by setting the 
derivative of $(4.30)$ to zero, and this gives rise to the 
equation 
$$\sum_{n=1}^{\infty}\frac {cos(\theta_R)cosh(2\pi Rn)-1}{(cosh(2
\pi Rn)-cos(\theta_R))^2}=0,\tag 4.33$$
which is not solvable.  However, we previously 
calculated the unique critical points for the terms in 
$(4.30)$, and from this we see that 
$$\theta_R\ge min\{\theta_{R,n}:n\ge 1\}=2\pi R.\tag 4.34$$
This will allow us to avoid multiple cases below.  

Since $cosh(x)\ge 1+x^2/2$, the function $(4.30)$ is bounded by 
$$2Re^{\pi R}sin(\theta )\sum_{n=1}^{\infty}\frac 1{(1-cos(\theta 
))+(2\pi Rn)^2}.\tag 4.35$$
The Poisson summation formula (applied to the function 
$f(x)=exp(-\vert x\vert )$) implies the identity 
$$\sum_{n=1}^{\infty}\frac 1{\alpha^2+n^2}=\frac 12(\frac {\pi}{\alpha}
cotanh(\pi\alpha )-\frac 1{\alpha^2}).\tag 4.36$$
This identity, with $\alpha^2=(1-cos(\theta ))/(2\pi R)$, implies that 
$(4.35)$ equals 
$$Re^{\pi R}sin(\theta )\frac 1{(2\pi R)^2}(\frac {\pi\cdot 2\pi 
R}{(1-cos(\theta ))^{1/2}}cotanh(\pi\alpha )-\frac {(2\pi R)^2}{1
-cos(\theta )})\tag 4.37$$
Thus $(4.30)$ is bounded by 
$$e^{\pi R}\frac {sin(\theta )}{2(1-cos(\theta ))^{1/2}}cotanh(\pi
\alpha ).\tag 4.38$$
For sufficiently small $R$, because of $(4.34)$, 
$$1-cos(\theta_R)\ge\frac 12(2\pi R)^2.\tag 4.39$$
This implies that $\alpha (\theta_R)\ge 1/$$\sqrt 2$.  Because $c
otanh$ is 
decreasing, and $sin(\theta )(1-cos(\theta ))^{-1/2}$ is bounded by $\sqrt 
2$, this 
implies that $(4.38)$ is bounded by 
$$e^{\pi R}2^{-1/2}cotanh(\pi /\sqrt 2)\le (.73)e^{\pi R}.\tag 4.40$$
This is bounded by $(4.32)$ (which is very close to $1$), for 
$R<1/10$ (using Maple).  // 

\smallskip

We now specialize to the case $l=0$.  In the analogue of 
$(3.7)$ we have 
$$\delta =\delta_R(r)=\frac 1{\Cal Z}(-\frac {\partial}{\partial 
r}\{\frac {\theta_3(Rr,iR)}{sinh(\pi /2R)^2+cosh(r)^2}\})sinh(r).\tag 4.41$$
This is positive, and its square root is the vacuum.  The 
corresponding potential function is given by 
$$q_R(r)=\frac {D^2(\delta^{1/2})}{\delta^{1/2}}=\frac 12[\frac {
\delta^{\prime\prime}}{\delta}-\frac 12(\frac {\delta'}{\delta})^
2]=\frac 12[(ln\delta )^{\prime\prime}+\frac 12(ln\delta )^{\prime 
2}].\tag 4.42$$

When $R=\infty$, this is $\frac 14-\frac {15}4sech^2(r)$, which is bounded.  
As we explained in the introduction, we would like to 
believe that the operator $-D^2+q_R$ has discrete spectrum, 
when $R<\infty$.  This is equivalent to showing that $q_R$ is 
unbounded as $r\uparrow\infty$, for $R<\infty$.  Unfortunately I have not 
been able to resolve this issue.  

\bigskip

\centerline{Appendix.}

\bigskip

We identify $\frak p$ with $\Bbb R^n$ (with $n=3$ in our rank one case), 
as in $(3.5)$ above, so that our preferred inner product on 
$\frak p$ is twice the Euclidean dot product.  Now suppose that 
we are given a metric on $\Bbb R^n$, 
$$g_x(\xi ,\eta )=A(x)\xi\cdot\eta ,\tag A.1$$
where $A(x)$ is a positive matrix for each $x\in \Bbb R^n$.  Let $
\nabla$, 
$dV$, ..  denote the usual Euclidean gradient, volume 
element, ...  

\proclaim{(A.2)Lemma}We have 

(a) $\nabla^g=A^{-1}\nabla$ 

(b) $dV_g=\rho dV$, $\rho =det(A)^{1/2}$ 

(c) $div^g(v)=\rho^{-1}div(\rho v)$, $v\in Vect(\Bbb R^n)$ 

(d) $\Delta^g=-\rho^{-1}div(\rho A^{-1}\nabla (\cdot ))$ 

(e) $Q_g(f)=\int (\Delta^gf)\bar {f}\rho dV=\int g(A^{-1}\nabla f
,A^{-1}\nabla f)\rho dV$.  

Assuming that $g$ is orthogonally invariant, we also have 

(f) For $f=f(r)$, 
$$Q_g(f)=\int_0^{\infty}f^{\prime 2}\alpha (r)\rho (r)r^{n-1}dr,\tag A.3$$
$$\Delta^g_r=-\delta^{-1/2}\circ D\circ\alpha\circ D\circ\delta^{
1/2}+\frac 12[\frac {(\alpha\delta')'}{\delta}-(ln\delta )^{\prime 
2}],\tag A.4$$
where $\delta (r)=\rho (r)r^{n-1}$ and $\alpha (r)=A^{-1}(x)\frac 
xr\cdot\frac xr$, for any 
choice of $x$ with $\vert x\vert =r$.  In particular, if $A$ is analytic 
and locally expressible as a power series in $ad(x)$, with 
$A(0)=1$, then $\alpha =1$.  
\endproclaim

\flushpar Proof of (A.2).  Parts (a)-(e) are routine, and the 
formula for $Q_g$ in (A.3) follows directly from (e).  For 
$(A.4)$, since $\Delta_r^g$ is self-adjoint with respect to $\delta 
(r)dr$, we 
know a priori that $\Delta^g_r$ has the form 
$$-\delta^{-1/2}\circ D\circ\alpha\circ D\circ\delta^{1/2}+\gamma 
.\tag A.5$$
We can plug this form into $Q_g$ in (e) and compare with 
$(A.3)$.  This determines $\alpha$ and $\gamma$.//  

\smallskip

\flushpar$(A.6)$Example.  Suppose first that we identify 
$\frak p\to G/K:x\to e^xK$, where the latter space is equipped with 
its negatively curved metric.  We have 
$$g^{G/K}_x(\xi ,\eta )=\langle\frac d{dt}\vert_{t=0}e^{x+t\xi},.
.\rangle_{e^x}$$
$$=A(ad(x))\xi\cdot\eta ,\tag A.7$$
where $A(x)=\vert\frac {1-e^{-z}}z\vert_{z=ad(x)}\vert^2$ (using a standard formula 
for derivative of the exponential map).  In the case 
$G=SL(2,\Bbb C)$, so that $\frak p=\Bbb R^3$, 
$$\rho (r)=det(\frac {1-e^{-z}}z\vert_{z=ad(x)})=\frac {1-e^{-r}}
r\cdot\frac {1-e^r}r$$
$$=\frac {2-2cosh(r)}{r^2}=\frac {sinh^2(\frac r2)}{(\frac r2)^2}\tag A.8$$
which is consistent with Harish-Chandra's formula, if we 
remember how things are normalized.  Now (f) implies 
that 
$$\Delta^{G/K}=-\delta^{1/2}\circ D^2\circ\Delta^{1/2}+\frac 14.\tag A.9$$
\smallskip

\flushpar(A.10)Example.  We consider the Guillemin-Stenzel 
Kahler structure for $G=K\times \frak p=K\times \frak k=TK$, where 
$\frak k\to \frak p:v\to iv$ ([St]).  This is interesting to consider, because 
there is a conjectural generalization of this to a general 
compact Riemannian manifold $X$ with $Ric\ge 0$ (a condition 
related to the renormalizability of sigma models).  

The complex structure is the usual one for $G$.  If we 
identify the tangent space of $K\times \frak p$ with $\frak k\otimes 
\frak p$ using left 
translation, then at the point $(k,x)\in K\times \frak p$, the complex 
structure is given by 
$$J_{(k,x)}(\left(\matrix \xi\\
\eta\endmatrix \right))=i\left(\matrix \frac {1-cosh(z)}{sinh(z)}&\frac {
2(cosh(z)-1)}{zsinh(z)}\\
\frac z{sinh(z)}&\frac {cosh(z)-1}{sinh(z)}\endmatrix \right)\left
(\matrix \xi\\
\eta\endmatrix \right)\vert_{z=ad(x)}\tag A.11$$
where ``$i$'' stands for usual multiplication by $i$ on 
$\frak g=\frak k\oplus \frak p$.  

The canonical $T^{*}K$ symplectic structure, in the Cartan 
coordinates $K\times \frak p$, is constant and given by 
$$\omega (\left(\matrix \xi\\
\eta\endmatrix \right),\left(\matrix \xi'\\
\eta'\endmatrix \right))=\langle i\xi\otimes\eta'-\eta\otimes i\xi'
\rangle ;\tag A.12$$
note that $i\xi\in \frak p$, so that the inner product makes sense.  

It follows from these calculations that the Riemannian 
metric is given by 
$$g_{(k,x)}(\left(\matrix \xi\\
\eta\endmatrix \right),\left(\matrix \xi\\
\eta\endmatrix \right))=\omega (J_{(k,x)}\left(\matrix \xi\\
\eta\endmatrix \right),\left(\matrix \xi\\
\eta\endmatrix \right))$$
$$=\omega (i\left(\matrix \frac {1-cosh(z)}{sinh(z)}&\frac {2(cos
h(z)-1)}{zsinh(z)}\\
\frac z{sinh(z)}&\frac {cosh(z)-1}{sinh(z)}\endmatrix \right)\left
(\matrix \xi\\
\eta\endmatrix \right)\vert_{z=ad(x)},\left(\matrix \xi\\
\eta\endmatrix \right))$$
$$=\langle (\frac {1-cosh(z)}{sinh(z)}\xi +\frac {2(cosh(z)-1)}{s
inh(z)}\eta )\otimes\eta -(\frac z{sinh(z)}\xi +\frac {cosh(z)-1}{
sinh(z)}\eta )\otimes\xi\rangle\vert_{z=ad(x)}$$
$$=\langle\frac {ad(x)}{sinh(ad(x))}\xi\otimes\xi +2\frac {cosh(a
d(x))-1}{ad(x)sinh(ad(x))}\eta\otimes\eta\rangle\tag A.13$$
where the bracket denotes the negative of the 
(appropriate multiple of) the Killing form.  

Now suppose that we just consider $\frak p$.  In this case 
$$A(ad(x))=2\frac {cosh(z)-1}{zsinh(z)}\vert_{z=adx}=\frac {tanh(
w)}w\vert_{w=\frac 12adx}.\tag A.14$$
and $\rho (r)=2tanh(r/2)/r$.  Again, $\alpha =1$.  

\bigskip

\flushpar Acknowledgement.  I thank Hermann Flaschka 
and John Palmer for helpful conversations.  

\bigskip

\centerline{References}

\bigskip

\flushpar[CG] K.  Clancy and I.  Gohberg, Factorization of 
Matrix Functions of Singular Integral Operators, 
Birkhauser (1981).  
 
\flushpar[Ga] K.  Gawedzki, Introduction to CFT, Quantum 
Fields and Strings:  A Course for Mathematicians, Vol.  
2, AMS-IAS, (1998).  

\flushpar[Gr] L.  Gross, Uniqueness of ground states for 
Schrodinger operators over loop groups, J.  Funct.  Anal.  
112 (1993), 373-441.  

\flushpar[Helg] S.  Helgason, Groups and Geometric 
Analysis, Academic Press (1984).  

\flushpar[L] G.  Lamb, Elements of Soliton Theory, John 
Wiley (1980).  

\flushpar[Pi] D.  Pickrell, Invariant measures for unitary 
forms of Kac-Moody Lie groups, Memoirs of the AMS, 
Vol 146, No 693 (2000).  

\flushpar[PS] A.  Pressley and G.  Segal, Loop Groups, 
Oxford University Press (1986).  

\flushpar[Sm] F.A.  Smirnov, Space of local fields in 
integrable field theory and deformed abelian differentials, 
Proceedings of the International Congress of 
Mathematicians, Berlin 1998, Doc.  Math.  J.  DMV,  Vol.  
III, 183-192.  

\flushpar[St] M.  Stenzel, Kaehler structures on cotangent 
bundles of real analytic Riemannian manifolds, Ph.D.  
thesis, M.I.T.  (1990).  

\flushpar[Wi1] E.  Witten, Supersymmetry and Morse 
theory, J.  Diff.  Geom.  17 (1982) 661-692.  

\flushpar[Wi2] ---------, Dynamics of quantum field 
theory, Quantum Fields and Strings:  A Course for 
Mathematicians, Vol.  II, AMS-IAS (1998).  

\flushpar[WW] E.  Whitaker and G.  Watson, A Course of 
Modern Analysis, Cambridge University Press, NY (1943).  

\flushpar[Z] A.  Zamolodchikov, Factorized S-matrices in 
two dimensions as the exact solutions of certain 
relativisitic quantum field theory models, Annals of 
Physics 120 (1979) 253-291.  

\end